\documentclass[12pt]{article}
\usepackage{graphicx}
\title{\bf Non-thermal X-ray and Gamma-ray Emission from the Colliding Wind Binary WR140}
\author{J.~M.~Pittard $^1$ and S.~M.~Dougherty $^2$\\
\vspace{1cm}\\
\normalsize $^1$School of Physics and Astronomy, The University of Leeds, Leeds, UK\\
\normalsize $^2$National Research Council, Herzberg Institute of Astrophysics, DRAO, Penticton, Canada}
\date{\mbox{}}
\begin{document}
\maketitle
\pagestyle{empty}
%
%
\def\bull{\vrule height .9ex width .8ex depth -.1ex}
\makeatletter
\def\ps@plain{\let\@mkboth\gobbletwo
\def\@oddhead{}\def\@oddfoot{\hfil\tiny\bull\quad
``Massive Stars and High-Energy Emission in OB Associations''; JENAM
   2005, held in Li\`ege (Belgium)\quad\bull}%
\def\@evenhead{}\let\@evenfoot\@oddfoot}
\makeatother
%
%
\def\beginrefer{\section*{References}%
\begin{quotation}\mbox{}\par}
\def\refer#1\par{{\setlength{\parindent}{-\leftmargin}\indent#1\par}}
\def\endrefer{\end{quotation}}
%
%
{\noindent\small{\bf Abstract:} 
WR\thinspace140 is the archetype long-period colliding wind binary
(CWB) system, and is well known for
dramatic variations in its synchrotron emission during its 7.9-yr,
highly eccentric orbit. This emission is thought to arise from
relativistic electrons accelerated at the global shocks bounding the 
wind-collision
region (WCR). The presence of non-thermal electrons and ions should also
give rise to X-ray and $\gamma$-ray emission from several
separate mechanisms, including inverse-Compton cooling, relativistic
bremsstrahlung, and pion decay. We describe new calculations
of this emission and make some preliminary predictions for the 
new generation of gamma-ray observatories. We determine that 
WR\thinspace140 will likely require several Megaseconds of
observation before detection with INTEGRAL, but should be a 
reasonably strong source for GLAST.}
%
%
\section{Introduction}
CWB systems are an important laboratory for investigating the
underlying physics of particle acceleration since they provide access
to higher mass, radiation and magnetic field energy densities than in
supernova remnants, which have been widely used for such
work. High-resolution observations (e.g., Williams et
al. 1997; Dougherty et al. 2000, 2005) show that the particle acceleration site
is where the massive stellar winds collide - the WCR. In addition to
synchrotron radio emission, these
non-thermal particles should give rise to X-ray and $\gamma$-ray
emission from several separate mechanisms, including inverse-Compton (IC)
cooling, relativistic bremsstrahlung, and pion decay, though
definitive evidence for such emission does not yet exist. The advent
of new and forthcoming observatories (INTEGRAL, GLAST and VERITAS)
will dramatically improve the chance of detecting these systems, and
in this paper we make new predictions for the expected flux from
WR\thinspace140.

Previous predictions of the IC emission have been based on
the assumption that the ratio of the luminosity from IC scattering to
the synchrotron luminosity is equal to the ratio of the photon energy
density, $U_{\rm ph}$, to the magnetic field energy density, $U_{\rm
B}$:
\begin{equation}
\label{eq:icsynclum}
\frac{L_{\rm ic}}{L_{\rm sync}} = \frac{U_{\rm ph}}{U_{\rm B}}. 
\end{equation}
However, the use of this equation is rather unsatisfactory for a
number of reasons. For example, $L_{\rm sync}$ is typically set to the {\em
observed} synchrotron luminosity, whereas free-free absorption by the
extended wind envelopes can be significant (see Pittard et al. 2005). 
In such cases the intrinsic synchrotron
luminosity, and thus the non-thermal X-ray and $\gamma$-ray luminosity,
will be underestimated. The predictive power of Eq.~1 is also greatly
undermined by the fact that the magnetic field strength in the WCR of
CWB systems is generally not known with any certainty. Since 
$U_{\rm B} \propto B^{2}$, small changes in the estimated value of $B$
can lead to large changes in $L_{\rm ic}$ (see Fig.~3 in Benaglia 
\& Romero 2003). 

A more robust estimate of the IC emission from CWB systems can be
acquired from model fits to radio data, where the population and
spatial distribution of non-thermal electrons is determined (see
Dougherty et al. 2003 and Pittard et al. 2005).  A key improvement in
these papers over earlier modelling of the radio emission from CWB
systems is the use of a hydrodynamical model to describe the stellar
winds and WCR. The free-free emission and absorption coefficients are
determined from the local temperature and density values on the
hydrodynamic grid. Particle acceleration at the shocks bounding the
WCR creates a population of non-thermal particles with a power-law
energy distribution (i.e. $n(\gamma) \propto \gamma^{-p}$, where
$\gamma$ is the Lorentz factor) - as these advect downstream IC
cooling modifies the energy distribution of non-thermal
electrons. Since the population of non-thermal particles is not
determined from first principles, their energy density is normalized
to some fraction ($\zeta_{\rm rel,e}$ and $\zeta_{\rm rel,i}$ for the
electrons and ions, respectively) of the thermal energy density
($U_{\rm th}$) at the shocks. For the non-thermal electrons, IC
cooling reduces this fraction below $\zeta_{\rm rel,e}$ in the
downstream flow. Similarly, the magnetic field energy density is
normalized by setting $U_{\rm B} = \zeta_{\rm B} U_{\rm th}$. When
modelling specific systems, $\zeta_{\rm B}$ and $\zeta_{\rm rel,e}$
are chosen to best match the observed radio emission.  Predictions for
the IC and relativistic bremsstrahlung emission then follow since
these arise from the same non-thermal relativistic electrons which are
responsible for the synchrotron radio emission.

\section{The High Energy Non-Thermal Emission}
In our current calculations we assume that IC scattering is isotropic
and takes place in the Thomson regime. Since the average photon energy
of an early-type star, $h\nu_{*} \sim 10\;$eV, Lorentz factors of
order $10-10^{4}$ are sufficient to produce IC X-ray and $\gamma$-ray
radiation. The resulting emission has a spectral shape which is
identical to the synchrotron emission at radio
frequencies. Bremsstrahlung radiation at $\gamma$-ray energies is
emitted from relativistic electrons since photons of comparable energy
to that of the emitting particle can be produced.  Finally, it is
possible to obtain $\gamma$-ray emission from hadronic collisions
involving non-thermal ions, through the decay of neutral pions, e.g.,
$p + p \rightarrow \pi^{0} + X$, $\pi^{0} \rightarrow \gamma +
\gamma$.  The pion decay process yields information on the population
of non-thermal nucleons, in contrast to the IC and bremsstrahlung
processes where the emission is dependent on the population of
non-thermal electrons. We conservatively set the energy density ratio
of non-thermal ions to electrons, $\zeta_{\rm rel,i}/\zeta_{\rm
rel,e}$, to 20.  Further background to the calculations can be found
in Pittard \& Dougherty (2006).

\section{Models of WR\thinspace140}
We apply our model to WR\thinspace140, the archetype of long-period
CWB systems. It consists of a WC7 star and an O4-5 star in a highly
elliptical orbit ($e \approx 0.88$), and in the context of the present
paper is most noteable for its possible association with an EGRET
source, lying on the outskirts of the positional error box of
3EG~J2022+4317 (Romero et al. 1999). We adopt the orbital parameters and
mass-loss rates determined by Dougherty et al. (2005), but allow the
mass-loss rate of the O star (and thus the wind momentum ratio $\eta$) 
to vary, as this parameter is not well determined. The system distance is
assumed to be 1.85~kpc. Further details on the modelling can be found in
Pittard \& Dougherty (2006) and Dougherty et al. (these proceedings).
  
Given the wealth of observational data which now exists for
WR\thinspace140, we have concentrated on obtaining a reasonable
spectral fit to the radio data at a specific orbital phase, namely
$\phi= 0.9$. Fig.~1 shows the resulting fit where it is clear that the
intrinsic synchrotron emission suffers significant free-free absorption by the
extended circumbinary envelope. A key point is that the large ratio
between the intrinsic and observed synchrotron emission implies that
the IC luminosity predicted using Eq.~\ref{eq:icsynclum} will be
substantially underestimated.

\begin{figure}
\centering
\includegraphics[width=16.0cm]{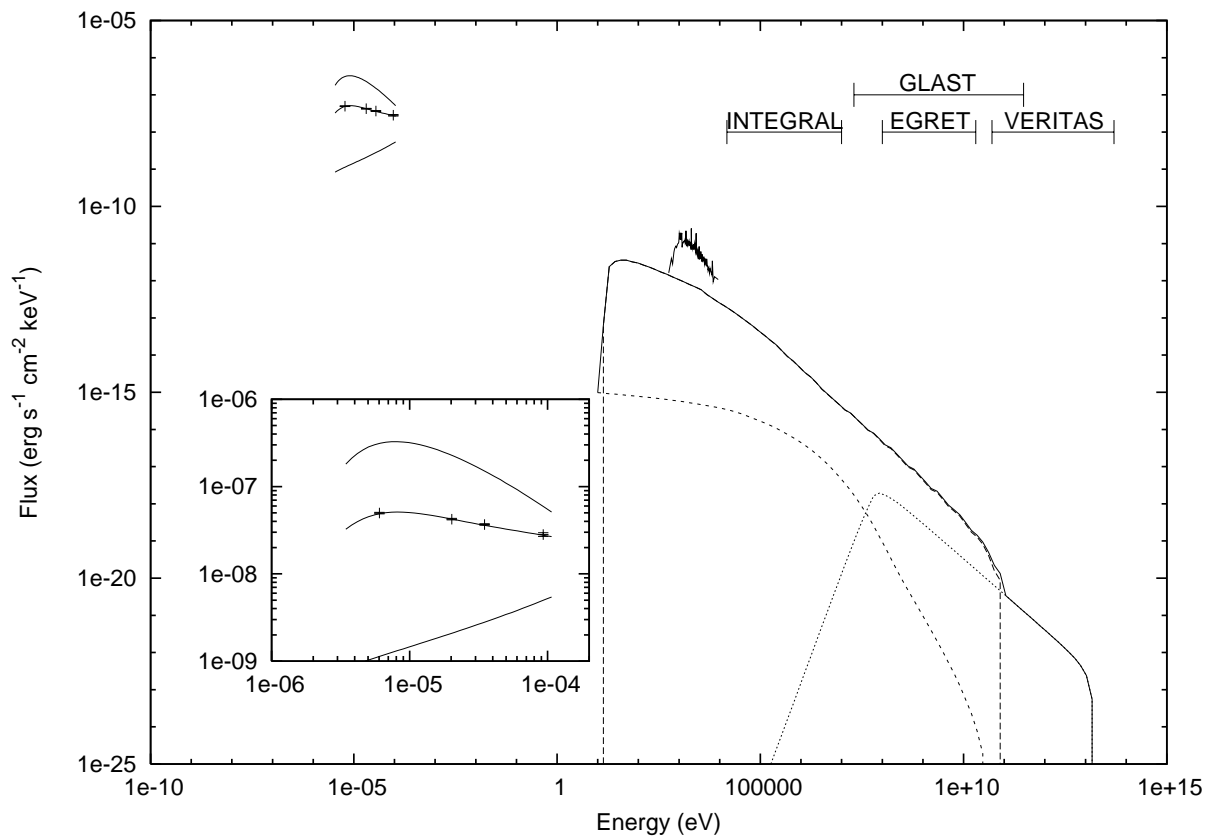}
\caption{The radio, and non-thermal UV, X-ray and $\gamma$-ray
emission calculated from our model of WR\thinspace140 at $\phi=0.9$,
together with the observed radio and X-ray flux. The model radio data
that we show indicate the thermal free-free flux (displayed below the
data points), the {\em intrinsic} synchrotron flux (displayed above
the data points), and the total observed emission. This region is
expanded in the inset.  The IC (long dash), relativistic
bremsstrahlung (short dash), and pion decay (dotted) emission are
shown as separate components, and their sum is also displayed
(solid). No absorption has been included in the calculation of the
high-energy emission.  We also show the observed X-ray spectrum
obtained at $\phi=0.84$ with {\it ASCA} (dataset 27022010, observed on
1999-10-22). It is reassuring that the IC flux predicted from our
model is less than the observed X-ray flux.}
\end{figure}

The high energy emission resulting from the radio fit is also shown
in Fig.~1. The IC emission dominates for photon energies less
than 50~GeV, while the emission from pion decay reaches energies up to
$15\;{\rm TeV}$. The relativistic bremsstrahlung emission is a minor
contributor to the total non-thermal flux, and does not dominate at
any energy. A gradual steepening of the IC spectrum occurs between
$10^{2}-10^{6}$~eV. The lack of a clearly defined spectral break reflects the
fact that the break frequency is spatially dependent, and so is smoothed
out once the flux is integerated over the entire WCR (Pittard et al. 2005).

The total photon flux in the 100~MeV - 20~GeV EGRET band from our
model is $3 \times 10^{-8} \;{\rm ph\;s^{-1}\;cm^{-2}}$, which is an
order of magnitude below the flux detected from 3EG~J2022+4317 during
the December 1992 observation at $\phi=0.97$.  The different phases
prevent a direct comparison, but 
we expect that the IC emission will
increase between $\phi=0.9-0.97$ as the stellar separation more than
halves. Considering the limitations of the present model, this level
of agreement is satisfactory. We predict a photon flux in the
15~keV-10~MeV sensitivity range of the IBIS instrument onboard
INTEGRAL of $1.4 \times 10^{-4} \;{\rm ph\;s^{-1}\;cm^{-2}}$, which is
likely to require an exposure time greater than 3~Msec for a $5\sigma$
detection.  The predicted 20~MeV-300~GeV photon flux is $1.4 \times
10^{-7} \;{\rm ph\;s^{-1}\;cm^{-2}}$, and thus WR\thinspace140 should
be fairly easily detected by GLAST. Our flux predictions are approximately
7 times lower than those made by Benaglia \& Romero (2003) - this
is partly due to the greater distance which we assume for WR\thinspace140.

\section{Conclusions}
We have presented predictions of the high energy non-thermal flux from
WR\thinspace140. We find that a long observation will be required
for detection of WR\thinspace140 with INTEGRAL, but it should be a 
reasonably strong source in
the GLAST energy band. The current neglect of two-photon pair
production prevents us from making predictions for the flux in the
sensitivity range of VERITAS, a Cherenkov air shower telescope array.
However, we expect that this absorption mechanism will significantly
attenuate the emission in the TeV range and suggest that other CWB
systems which are wider and less distant (e.g., WR\thinspace146 and
WR\thinspace147) may be better targets at these energies.  Improved
high energy data will drive the development of theoretical models which include
anisotropic IC emission and two-photon pair production, and will allow
a much more detailed comparison between observations and predictions.

%
%
\section*{Acknowledgements}
JMP gratefully acknowledges funding from the Royal Society.
%
%
 
\beginrefer
\refer Benaglia P., Romero G.E., 2003, A\&A, 399, 1121

\refer Dougherty S.M., Beasley A.J., Claussen M.J., Zauderer B.A., 
Bolingbroke N.J., 2005, ApJ, 623, 447

\refer Dougherty S.M., Pittard J.M., Kasian L., Coker R.F., Williams P.M.,
Lloyd H.M., 2003, A\&A, 409, 217

\refer Dougherty S.M., Williams P.M., Pollacco D.L., 2000, MNRAS, 316, 143

\refer Pittard J.M., Dougherty S.M., Coker R.F., Bolingbroke N.J., 
O'Connor E., 2005, A\&A, submitted

\refer Pittard J.M., Dougherty S.M., 2006, MNRAS, submitted

\refer Romero G.E., Benaglia P., Torres D.F., 1999, A\&A, 348, 868

\refer Williams P.M., Dougherty S.M., Davis R.J., van der Hucht K.A., 
Bode M.F., Setia Gunawan D.Y.A., 1997, MNRAS, 289, 10

\endrefer           
\end{document}